# DISTORTION IN OPHTHALMIC OPTICS:
# A REVIEW OF THE PRINCIPAL CONCEPTS AND MODELS

Jean-Marie Hanssens OD PhD [a]
Bernard Bourdoncle Ing [b]
Jacques Gresset* OD PhD [a]
Jocelyn Faubert PhD [a]
Pierre Simonet* OD PhD [a]

a – Université de Montréal, École d'optométrie
CP 6128, Succursale Centre-Ville
Montréal, Québec, Canada.
*Emeritus professor

b – Essilor International, R&D Vision Science Director
57 rue de Condé
94106 Saint-Maur Cedex, France

**Corresponding authors:**

Name:     Jean-Marie Hanssens & Jocelyn Faubert
Address:  School of optometry - Université de Montréal
          C.P. 6128, Succursale Centre-Ville
          Montreal, QC, H3C 3J7. Canada
Phone:    514-343-8805
Fax:      514-343-2382
Email:    jean-marie.hanssens@umontreal.ca, Jocelyn.faubert@umontreal.ca

# 1 Abstract

Although all members of the ophthalmic community agree that distortion is an aberration affecting the geometry of an image produced by the periphery of an ophthalmic lens, there are several approaches for analyzing and quantifying this aberration. Various concepts have been introduced: ordinary distortion, stationary distortion and central static distortion are associated with a fixed eye behind the ophthalmic lens, whereas rotatory distortion, peripheral distortion, lateral static distortion, and dynamic distortion require a secondary position of gaze behind the lens. Furthermore, concept definitions vary from one author to another. The goal of this paper is to review the various concepts, analyze their effects on lens design and determine their ability to predict the deformation of an image as perceived by the lens wearer. These entities can be classified within 3 categories: the concepts associated with an ocular rotation, the concepts resulting from an optical approach, and the concepts using a perceptual approach. Among the various concepts reviewed, it appears that the Le Grand-Fry approach for analyzing and displaying distortion is preferable to others and allows modeling of the different possible types of distortions affecting the periphery of an ophthalmic lens.

## 2  Purpose

Among the off-axis aberrations affecting the optics of ophthalmic lenses, distortion alters the geometry of the image and not its sharpness as in the case of oblique astigmatism, mean oblique power error or transverse chromatic aberration. For this reason, lens designers as well as clinicians have not given distortion the full attention it deserves given it is a factor in modifying the visual perception of the lens wearer.

Even though the majority of authors initially described the qualitative visual effect caused by distortion, the quantitative evaluation of this aberration and its analysis in terms of the deformation of images perceived by the lens wearer has generated several approaches among the scientific community.[1-3] This article aims at defining the different concepts included within the umbrella term of "distortion", at analyzing the proposed methods for the correction or the reduction of this aberration, and at examining with a critical eye, in terms of the perception of the lens wearer, the accuracy of the concepts used.

A critical analysis of the scientific publications concerning distortion separates the term into three categories based on the different approaches developed to define and qualify this aberration. First, we will distinguish the concepts associated with ocular rotation, then that derived from the optical approach and, finally, the one related to a perceptual approach.

## 3  Approach associated with ocular rotation

Historically, Tscherning[3] was the first to analyze the phenomena of distortion induced by an ophthalmic lens and to attempt to correct this disorder:

> "Distortion is an error which often affects optical images: straight image lines set in the periphery of the field, are reproduced, in the image, by curves, which sometimes turn their concavity towards the center, (barrel deformation) and sometimes their convexity (pincushion deformation)".

Tscherning[3] stated:

"Distortion is due to the fact that the magnification varies with the distance from the axis. It occurs because the lens, which forms the image, is not aplanatic in relation to the distance at which the diaphragm is located".

In his calculations for correcting distortion, Tscherning[3] set the diaphragm in the center of rotation of the eye to obtain lenses without spherical aberration between this point and its conjugate point through the lens. The solution he obtained resulted in meniscus lenses with a high curvature, which he called orthoscopic lenses. He found for each power between +13 and − 20 diopters and for a center of rotation located 28 mm from the cornea, two forms of lenses which fulfil the conditions imposed by his calculations. In fact, these results were already published without demonstration since 1904, together with the results outlining the conditions for correction of oblique astigmatism. Tscherning[4] considered that lenses having the Wollaston form for the correction of oblique astigmatism were almost similar to the less curved orthoscopic lenses, which led him to produce and favor the use of such lenses.[5, 6]

It should be noted that in his initial work, Tscherning[4] did not use the term distortion, but orthoscopic error. Moreover, beside the correction of oblique astigmatism proposed in his publication, Tscherning[4] also tried to correct the mean oblique power error, but with no success since the solution he proposed was wrong. The same situation occurred for distortion, since Le Grand[7] indicated that Tscherning's[3, 4] approach does not prevent dynamic deformation of the image of a vertical line when the lens wearer moves his visual axis behind the lens from the optical center towards the periphery. Therefore, by choosing the center of rotation as a reference point in calculating distortion, Tscherning[3, 4] ruled out the analysis from perceptual and physiological basis, since other authors, following his example, will adopt an approach of distortion which will remain linked to ocular rotation.

Henker[2] and optical engineers from Zeiss suggested that the origin of distortion is the displacement of the conjugate of the center of rotation of the eye along the optical axis of the lens when the ocular rotation increases. According to Henker [2], the Airy condition (the relation between the angle of the ocular rotation behind the lens *u'*, and the angle of eccentricity in the field of fixation *u* (Figure 12), such as the ratio (tan *u'*/ tan *u*) remains constant, which is the condition for the absence of distortion in optical instruments) is not satisfied for ophthalmic lenses because of the displacement of the conjugate of the center of rotation of the eye through the lens as *u'* increases.

*Figure 1*

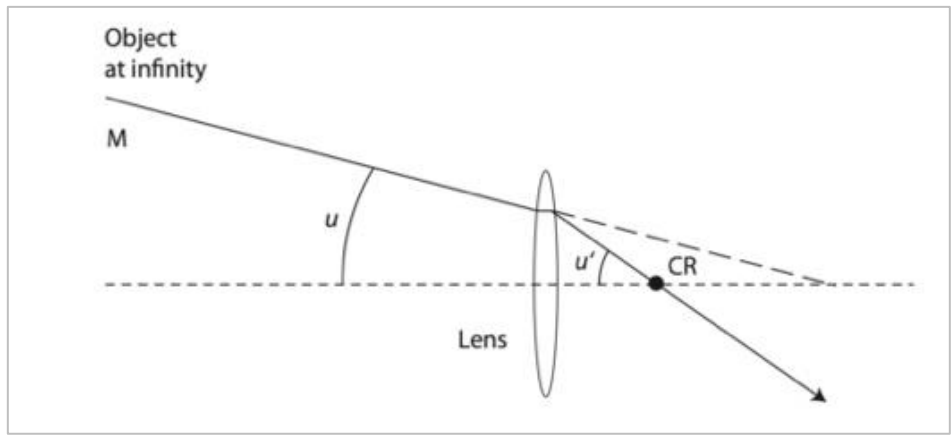

In the distortion linked to ocular rotation, the principal ray intersects the optical axis of the lens at the center of rotation of the eye (CR). The ocular rotation behind the lens is angle *u'* and angle *u* is the eccentricity of the fixation point M in the field of fixation.

Henker[2] evaluated the distortion as a percentage for a given angle of rotation *u'*, by considering the discrepancy between the angle *u*, assessed from the paraxial image of the center of the rotation, and the angle $u_r$, assessed from the actual position of the conjugate of the center of rotation given by refraction through the periphery of the lens. According to Bennett,[8, 9] Henker would have proposed the following equation for distortion:

Equation 1: D (%) = 100 (W-w)/W

with W = tan $u'$ / tan $u_r$ and w = tan $u'$ / tan $u$

In the aphakic patient, Henker[2] perfectly described the perception of barrel deformation of a vertical line observed during rotation of the head when the individual focuses his attention on the middle of the vertical line and performs an ocular rotation behind the lens in the opposite direction to the rotation of the head as his visual axis moves from the center towards the periphery of the lens. However, the concept of distortion quantified by Equation 1 does not correspond to the one associated with the description of the perceptual effect. In fact, Henker's[2] calculations are related only to the shift in the frontal plane, from a paraxial position, of the middle of the vertical line during the modification of ocular rotation. The distortion associated with the perception of curvature of the vertical line for the lens wearer, as the eye rotates behind the lens, results from the displacement of the extremities of the line relative to its central part, this aberration cannot be calculated according to the model proposed by Henker.[2]

Emsley[1, 10] investigated the distortion linked to rotation of the eye and considered the amplitude of ocular rotation when the eye explores the field of fixation. Emsley[1] called rotatory distortion, the variations of the ocular rotation when equal extents of the field of fixation are explored as the eye rotates. Emsley[1] calculated the area of the field of fixation (*du*) subtended by an ocular rotation of 1-degree (*du'*) and identified the ratio *du'/du* as "field angle magnification" (Figure 13). For a given base curve of the lens, the angular extent of the field of fixation (*du*) linked to a 1-degree rotation (*du'*) changes when the eye moves from the center to the periphery of the ophthalmic lens. For convex lenses, the ratio *du'/du* increases as the eye rotates. This change in the field angle magnification (*du'/du*) across the field of fixation represents rotary distortion, according to Emsley.[1]

*Figure 2*

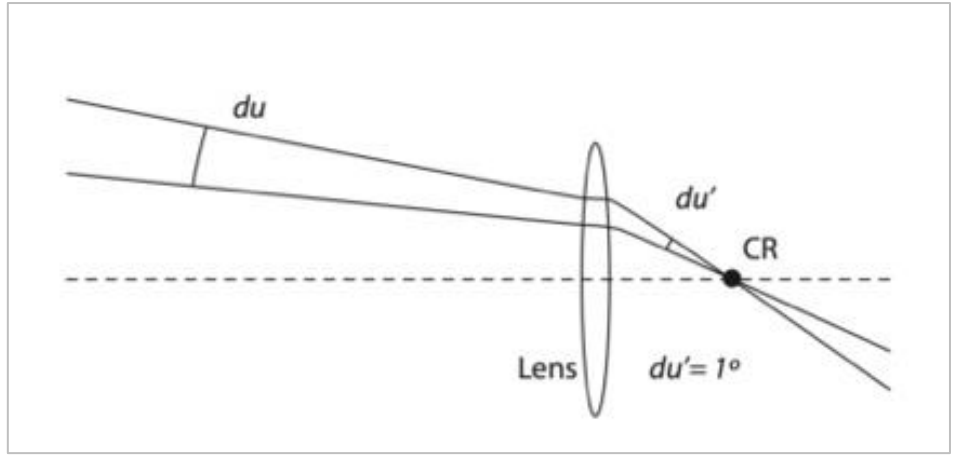

Rotatory distortion, according to Emsley[1], is associated with the change in the angular extent of the field of fixation (*du*) linked to 1 degree ocular rotation (*du'*) while the eye moves from the center to the periphery of the ophthalmic lenses.

Although, according to Le Grand[11], this concept of rotatory distortion is more related to the gauging linearity of the field of fixation than to the deformation of lines in the image, it would be of interest to know if this type of distortion can be corrected. The correction of rotatory distortion implies that the field angle magnification remains constant across the whole lens, even though the spherical aberration affects the conjugate of the center of rotation. By using third-order calculations, Emsley[10] obtained a relation between *u* and *u'*, when the following parameters of the lens are known: the lens power (F'), its refractive index (n) and the vergence of the center of rotation of the eye calculated with respect to back surface of the lens (Z). This relation is given by:

Equation 2:
$$u/u' = [(1-2F')/Z] - u'^2 \times (3.336/10^6) F' [24D_2^2 + D_2 (900 - 36F') + 18F'^2 - 675F' + 7105]$$

The ratio *u/u'* is a constant [(1-2F')/Z] if the rest of the relation is equal to zero. This condition corresponds to a quadratic equation for $D_2$ (power of the back surface of the lens), and so it is directly related to the curvature of the lens. This indicates that the correction of rotatory distortion is possible: there are two base curves which allow the ratio *u/u'* to be constant for lens powers between –21.00 D and +10.00 D. According to Emsley,[10] this quadratic equation, which modifies the constancy of the ratio *u/u'*, would constitute the expression of rotatory distortion. Emsley[1, 10] was also looking into the conditions leading to correction of rotatory distortion through exact trigonometrical ray tracing. Depending on the refractive power, the base curve of the lens required for correction of this type of distortion would be steeper than that required for the correction of oblique astigmatism when vision is at infinity. The difference is 2 diopters for a refractive power of +10 D, and reaches 10 diopters for a power of –21 D.

Bennett[8, 9] viewed distortion quite differently from Emsley,[1] because he proposed a quantitative evaluation of this aberration inspired by the approach advocated by Henker.[2] Like the previous German lens designers, Bennett called w, the ratio between the tangent of two angles, (*u'*) the angle of ocular rotation and ($u_o$) the angle of eccentricity in the field of fixation, this latter angle being calculated from the paraxial conjugate of the center of rotation of the eye. Bennett[8] considered the ratio w as analogous to the paraxial magnification existing at the entrance pupil, and he proposed to express it relative to the center of rotation of the eye, in paraxial conditions:

$$w = \tan u' / \tan u_o = \{1/ [1- t/n)D_1]\} [1/ (1- zF')]$$

with z corresponding to the radius of the vertex sphere, $D_1$ to the power of the front surface and t to the central thickness of the lens.

On the other hand, Bennett[8] defined the quantity W (W = tan *u'* / tan $u_1$), as the ratio between the tangent of the angle of the ocular rotation *(u')* and the tangent of the real angle of eccentricity in the field of fixation ($u_1$). The angle $u_1$ is calculated relative to the conjugate of the center of rotation of the eye, whose exact position is calculated trigonometrically by ray tracing, considering the spherical aberration linked to ocular rotation. Bennett[8] suggested expressing the distortion in percentage by an equation which differs from Henker's:

$$\text{Equation 3: } D\,(\%) = 100\,[(W - w)/w]$$

For an ocular rotation of 35° and for lenses of +6 and –6 D, he showed that distortion, as expressed from Equation 3, decreases if the base curve of the lens becomes steeper. Bennett[9] suggested that the minimum distortion of an ophthalmic lens for a given ocular rotation is observed when the effect of deviation associated with refraction of the lens periphery is divided equally between the two sides of the lens. He deduced by approximation, based on Prentice's formula, the condition that has to be fulfilled by the front surface curve of the lens ($D_1$), that is:

$$D_1 = [(2 - n)/2]F' + (n-1)Z$$

This corresponds to the equation of a straight line tangential to the Wollaston branch of Tscherning's ellipse at the level of a zero-power lens. Therefore, minimum distortion could be obtained for lenses that are highly curved such that, for a vertex sphere of 27 mm and a refractive index of 1.5, $D_1$ would be approximately +19.75 D and +17.25 D for lenses of +5 and –5 D respectively.

The criticisms directed towards Henker's approach also apply to Bennett's approach[8, 9], since this concept of distortion, and the quantification of this aberration, do not correspond to the lens wearer's perception of image distortion when he explores the lens periphery.

Bennett and Edgar[12] correctly concluded that all questions concerning distortion caused by ophthalmic lenses were not solved because certain important factors arise from the domain of visual perception. In the absence of experimental data relating to these factors, Bennett and Edgar[12] formulated the following assumptions, as a logical approach from the point of view of lens designers: on one hand, the eye aims to have the advantage of foveal vision; consequently, it scans the image it observes. On the other hand, the homogenous geometrical transformation that scales similarity between the object and the image should have, as a center of reference, the center of ocular rotation to consider the eye and the ophthalmic lens in conjunction.

With such premises, Bennett and Edgar[12] considered the apparent position of an object seen through the lens after an ocular rotation (*u'*), while the object would be seen by the naked eye with an ocular rotation (*$u_o$*) in a frontal plane. Thus, the parameter (*$u_o$*) represents a different quantity than the one identified with the same symbol by Bennett.[8] Bennett and Edgar[12] defined the quantity W as the *ocular rotation factor*, as such: W = tan *u'* / tan *$u_o$* (Figure 14). According to Bennett and Edgar,[13] distortion results from the fact that this ocular rotation factor is not constant in all directions of gaze, and they compute this aberration using equation 3 and expressing the distortion in percentage for a given ocular rotation, generally 30°. In this equation, w represents the ratio tan *u'* / tan *$u_o$* obtained when *u'* is low and when the paraxial calculation can be used.

*Figure 3*

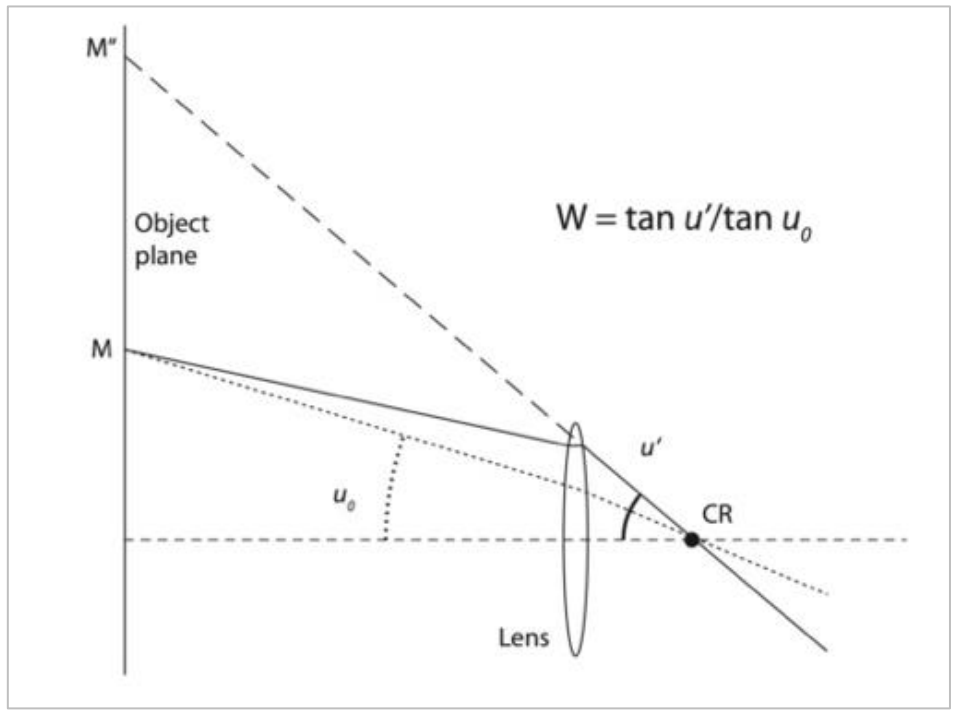

The ocular rotation factor W (Bennett and Edgar[12, 13]) is the ratio $\tan u' / \tan u_o$. M" is the apparent position of an object M seen through the lens after an ocular rotation $u'$ while the object M is seen by the naked eye with an ocular rotation equal to $u_o$. According to these authors, distortion results from the fact that W is not constant in all directions of gaze.

According to Bennett and Edgar.[13]

Equation 4: $w = \{L_2[1-(t+z)L_1]\} / [L'_1(1-zL'_2)]$

$L_1$, being the vergence of the object relative to the front surface

$L_2$, being the vergence of the object relative to the back surface

$L'_1$, being the vergence of the image relative to the front surface

$L'_2$, being the vergence of the image relative to the back surface

t, being the central thickness

z, being the radius of the vertex sphere

and

$$\text{Equation 5: } W = \{[1-(t+z)L_1] \tan u'\} / [-y_1 L_1 + (1 + x_1 L_1) \tan u]$$

With *u*, angular eccentricity of the object in the field of fixation given by the lens, while $x_1$ and $y_1$ are the co-ordinates of the incidence point on the front surface of the lens, the origin of the co-ordinates being at the apex of the lens. In the case when the object is at infinity, Equation 4 and 5 become respectively:

$$w = \{1/[1 - (e/t)D_1]\} [1/(1 - zF')]$$
$$\text{and } W = \tan u' / \tan u$$

Bennett and Edgar[12, 13], provided the results of their calculations of distortion for powers between +8 and −24 D, for *u'* = 30 degree and z = 27 mm for lenses with equi, plano or meniscus shapes. For these results, the authors also calculated the minimal value of this type of distortion and the base curve required in order to obtain this minimal distortion. It is interesting to note that these base curves are substantially different than the ones proposed by Emsley.[1] However, they remain quite similar to the ones predicted by Bennett.[9]

In conclusion, authors such as Tscherning,[3, 4] who associated distortion with ocular rotation, adopt an approach which does not allow the definition and quantification of distortion in terms of image deformation in the same way a lens wearer perceives it. Furthermore, Le Grand[7] does not see the theoretical nor practical interest of keeping the field angle magnification constant as in Emsley's[1] approach of rotatory distortion. As for the approach of Henker[2], Bennett[8, 9] or Bennett and Edgar[12, 13], the distortion they defined is based on the comparison of two ocular rotation factors, one for a position of gaze remaining in the paraxial domain, and the other for a secondary position of gaze through the periphery of the ophthalmic lens. The validity of this concept is questionable because the subjective perception of the deformation of a line in the lens periphery is not linked to the comparison of the two conditions, but rather to the direct observation of an image. The premises proposed by Bennett and Edgar[12] for the definition of distortion are certainly not the ones used by the ametrope as he perceives the deformations of the image caused by this aberration. However, Bennett and Edgar[12] recognized that the distortion caused by ophthalmic lenses involved perceptual factors.

## 4 Optical approach to distortion

Emsley[1] was the first to consider the distortion affecting a stationary eye, distinguishing it from distortion associated to an ocular rotation. Emsley[1] called *ordinary distortion* the distortion existing when the eye is fixed and when the entrance pupil of the eye, rather than the center of rotation of the eye, forms the aperture stop for the ophthalmic lens.

This type of distortion corresponds exactly to the classical approach in optical instrumentation. Therefore, if the object is at infinity, distortion is defined in the plane of the paraxial secondary focal point and expressed by the difference of position between the principal ray passing, after refraction, through the center of the entrance pupil and a paraxial ray passing through the nodal points of the spectacle lens with the incidence, $u$. This incidence corresponds to the eccentricity of the object within the visual field of the corrected ametrope (Figure 15).

*Figure 4*

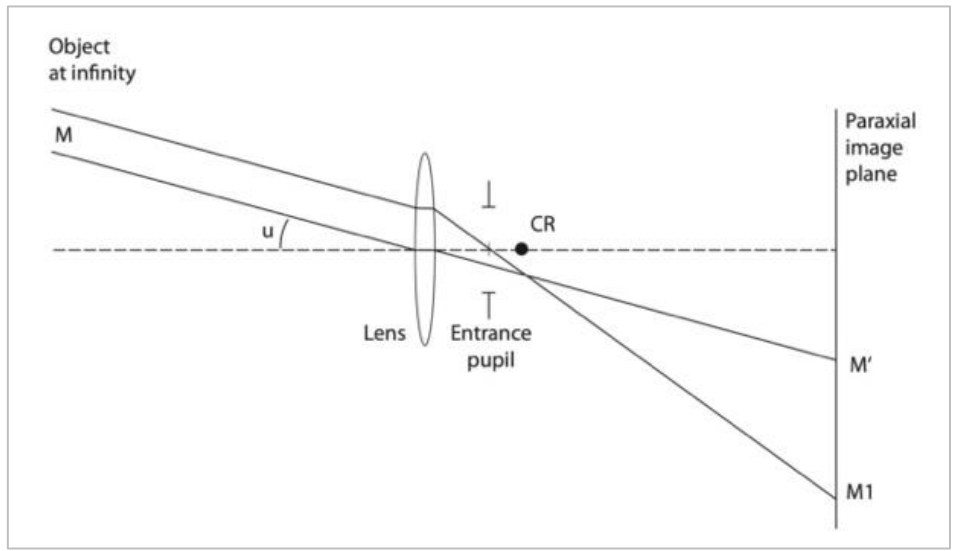

In ordinary distortion Emsley[1]; Jalie[14]; Katz[15]) or in stationary distortion (Atchison and Smith[16]), the principal ray, parallel to the paraxial ray having an incidence *u*, intersects the optical axis of the lens, after refraction, at the center of the entrance pupil of the eye. Distortion is quantified by M'M1, the difference of position in the plane of the paraxial secondary focal point between the principal ray and the paraxial ray. This type of distortion corresponds to the classical approach in optical instrumentation.

Under these conditions, the correction of this type of distortion requires that Airy's condition is fulfilled, and the pupil is free of spherical aberrations, which seems to be impossible for an ophthalmic lens with regards to the position of the pupil with respect to the lens. However, this type of distortion may be minimized by selecting the base curve of the lens for which the power, the refractive index, and the position with respect to the pupil are previously known. For a 30 degree ocular rotation, Emsley[1] determined the base curve by using trigonometrically ray tracing for lenses of +4.00 D and –5.00 D. The values obtained are too high to be realistic since the lens of +4.00 D should have had a back surface of –27.00 D., whereas for the –5.00 D lens, we would have found a value of –41.00 D. Within the interval from –20.00 D to +22.00 D, Emsley[1] used the third-order approximation to determine the curvature of the back surface of lenses minimizing this type of distortion. The curvature values varied from –49.00 D for a lens of –20.00 D to –20.00 D for a lens of +2.00 D, whereas an afocal lens should have had a back surface of –35.75 D. Thus, Emsley[1] has evaluated the correction of this ordinary distortion as impractical.

Jalie[14] used the concept of ordinary distortion proposed by Emsley[1] to define the distortion observed when the center of the entrance pupil is considered as a reference stop, however, he also introduced a concept of rotatory distortion that, however, differed from Emsley's. Jalie[14] applied the concept of distortion previously used for the center of the entrance pupil to the center of rotation of the eye. For Jalie[14], rotatory distortion represented, in the paraxial image plane, the difference in position between the paraxial image and the real image when the center of rotation is the reference stop (Figure 16).

*Figure 5*

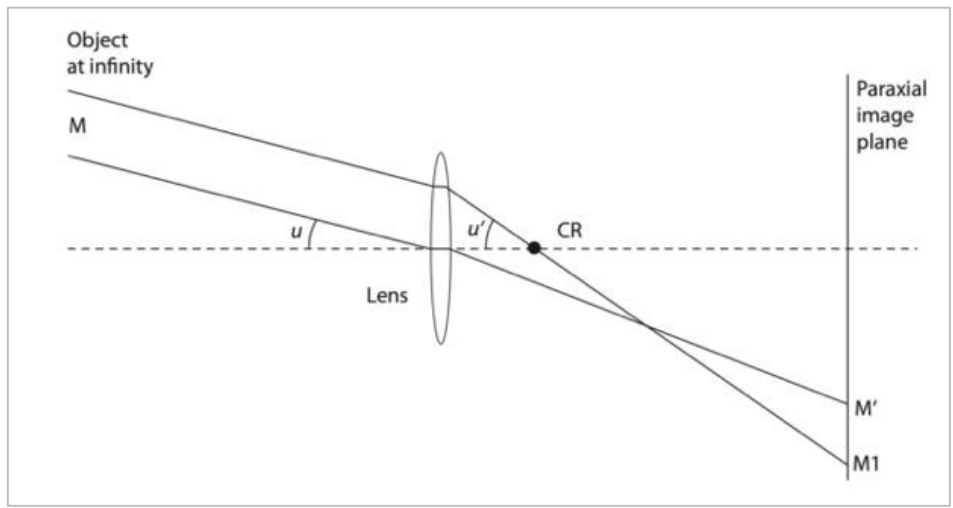

In rotatory distortion, according to Jalie[14] and to Atchison and Smith[16], the principal ray, parallel to the paraxial ray having an incidence *u*, intersects the optical axis of the lens, after refraction, at the center of rotation of the eye (CR). Distortion is quantified by M'M1, the difference of position in the plane of the paraxial secondary focal point between the principal ray and the paraxial ray.

Jalie[14] demonstrated that cancelling rotatory distortion requires that:

$$\tan u' / \tan u = \{1/[1 - (t/n)D_1]\} [1/(1 - zF')]$$

This condition is only possible within the paraxial approximation, that is when the values *u'* and *u* in Figure 16 are small. Jalie[14] expressed rotatory distortion as percentage rates:

$$D (\%) = 100 [(W - w)/w]$$

where W and w equal the ratio tan *u'* / tan *u* respectively for a given ocular rotation and for the paraxial condition. Thus, rotatory distortion, as defined by Jalie[14], is like the variation of the ocular rotation factor proposed by Bennett and Edgar.[12] Jalie[14] has calculated rotatory distortion for a lens of +10.00 D and has demonstrated that this aberration cannot be corrected. However, it may be minimized with adequate lens curvature. In order to derive the equation for finding the optimal curvature as a function of lens power, Jalie[14] adopted the same approach as Bennett.[9]

Katz[15] maintained the same approach as Emsley[1] concerning ordinary distortion but introduced a new concept about rotatory distortion, as defined by Jalie.[14] He proposed the evaluation of this aberration by considering both the intersections of the paraxial ray and of the ray passing through the center of rotation of the eye with the far-point sphere (Figure 17) and no longer with the paraxial secondary focal plane, that is a plane perpendicular to the optical axis and therefore perfectly flat.

*Figure 6*

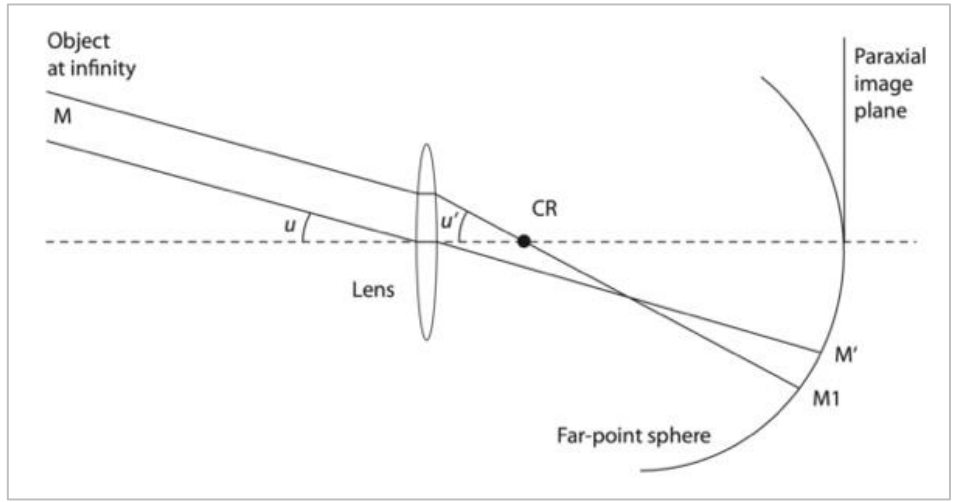

In rotatory distortion, according to Katz[15], the principal ray, parallel to the paraxial ray having an incidence *u*, intersects the optical axis of the lens, after refraction, at the center of rotation of the eye (CR). Distortion is quantified by M'M1, the difference of position on the far-point sphere between the principal ray and the paraxial ray.

In order to correct this type of rotatory distortion, the following condition is required:

$$\tan u' / \tan u = [f - r(1 - \cos u')] / r \cos u'$$

where f represents the secondary focal length measured from the principal plane and where r equals the radius of the far-point sphere. Katz[15] demonstrated that this condition is not satisfied for lenses corrected for oblique astigmatism. When rotatory distortion is evaluated on a flat focal plane, Katz[15] suggested the use of the following formula:

$$D(\%) = 100\,[(Y_f - y_f)/y_f]$$

where $Y_f$ and $y_f$ are the intersections of two rays with the paraxial secondary focal plane. According to Katz[15], this gives identical values to those obtained from equation 3.

When rotatory distortion is expressed with respect to the far-point sphere, Katz[15] suggests the following formula:

$$D\% = 100 \{[(Y_s - y_s) + (X_s - x_s)]^{1/2} / (y_s^2 + x_s^2)^{1/2}\}$$

where $X_s$, $Y_s$ and $x_s$, $y_s$ are respectively the coordinates of the intersections of the far-point sphere with the principal ray and the paraxial ray. The origin of these coordinates is at the secondary focal point, that is, the apex of the far-point sphere.

For lenses with powers varying between +6.60 D and –24.65 D and for which oblique astigmatism is corrected, Katz[15] demonstrated that rotatory distortion is less when calculated at the far-point sphere (variation from +0.85 to –13.54%) than when calculated for the paraxial secondary focal plane (variation from + 4.45 to –28.31%). Moreover, Katz[15] established that the minimum rotatory distortion for lenses, whose curvature was previously determined to minimize this effect, is shown to be less if the calculation is made in relation to the far-point sphere rather than to the flat secondary focal plane. According to Katz[15], the reduction of rotatory distortion occurring when the far-point sphere is used as a reference surface would explain why this type of distortion is hardly perceived by lens wearers.

Katz[17] put forward the importance of aspheric surfaces for negative powered lenses in order to correct simultaneously, rotatory distortion, oblique astigmatism and mean oblique power error, even though rotatory distortion was expressed in relation to a flat secondary focal plane. However, from his computation with aspheric surfaces, it appears that the lens still presents an excessively steep base curve, which is incompatible with conventional methods of surfacing or with their wear.

With third-order calculations, Atchison and Smith[16] have determined the distortion in thin lenses having an aspheric conicoidal surface. Atchison and Smith[16] called peripheral distortion or stationary distortion, the aberration in which the entrance pupil is considered as the reference stop. In fact, they are repeating Emsley's[1] conceptualization of ordinary distortion. They have also studied rotatory distortion defined in relation to the center of rotation of the eye and to a secondary focal paraxial plane, in a similar way to Jalie[14].

Atchison and Smith[16] indicated that both types of distortion obtained by third-order calculations cannot be corrected by using spherical surfaces and hence, they derived the curvature of the back surface of the lens in order to obtain the minimum distortion. They found a linear relationship that differs from those proposed by previous authors.[8]

By introducing to a lens an aspheric conicoidal surface, Atchison and Smith[16] demonstrated that the third-order expression for both types of distortion becomes a cubic function for the paraxial curvature of the lens. For a given power and asphericity, this signifies that there exists at least one lens curvature correcting each type of distortion. The calculations of Atchison and Smith[16] prove that, for a given asphericity, the correction of ordinary distortion requires a greater lens curvature than rotatory distortion. Furthermore, for a given asphericity, the paraxial curvature associated with the correction of rotatory distortion is steeper than the one required for the correction of oblique astigmatism. It remains possible to correct rotatory and oblique astigmatism simultaneously with an aspheric surface. However, the surface must have a paraxial base curve of very high value that renders its usage very unlikely. For example, a lens with a vertex power of –12.00 D and an asphericity Q of +0.51 D requires a paraxial back surface $D_2$ of –40.43 D, whereas a lens with a vertex power of +12.00 D and asphericity of –0.16 D would require a back surface of –18.83 D.

According to Atchison and Smith[16], the simultaneous correction of oblique astigmatism and stationary distortion is theoretically possible in the third-order approximation if conicoidal aspheric surfaces are employed, but this correction requires a steeper lens curvature than the one used to correct rotatory distortion and oblique astigmatism simultaneously. Atchison and Smith[16] have demonstrated that the correction of both types of distortion in lenses which present high powers and curvatures acceptable from a cosmetic point of view, leads to pronounced errors in the peripheral power. In the case of high-power convex lenses, asphericities required to correct power aberrations within the periphery of the lens tend to significantly decrease rotatory distortion but have no effect on stationary distortion.

In conclusion, the optical approach to distortion stems from instrumental optics and compares, once again, two images. One is a paraxial image and the other is an image given by refraction through the periphery of the lens. This approach remains valid from a theoretical optics standpoint, if one considers an ophthalmic lens as an independent optical system and does not refer to a lens used in combination with both the eye and the visual system. Consequently, it is far from clear if this is an appropriate mode of analysis and quantification for evaluating distortion as perceived by an ophthalmic lens wearer.

## 5 Perceptual approach to distortion

The approach adopted by Le Grand[11] can be considered original and interesting because the distortion introduced by ophthalmic lenses is analyzed in terms of the perception of the wearer and in terms of the actual deformation of images. Le Grand's[11] work was published in French and remained ignored.

Le Grand[11] distinguished three types of distortion. First, central static distortion is the aberration affecting the eye in primary position of gaze when the lens wearer's visual axis is superimposed with the optical axis of the ophthalmic lens during a fixation at the center of the frontal plane (Figure 18). In this case, the image of a vertical line AB located in the periphery of the frontal plane is given by refraction through the edge of the ophthalmic lens, and the center of the entrance pupil of the eye (P) is the point on which the rays originating from the vertical line lie after being refracted by the periphery of the ophthalmic lens (Figure 18). The orientation, at the center of the entrance pupil, of a refracted ray originating from point A of the vertical line gives the direction along which the image of this point is projected in the frontal plane (A"). In the presence of distortion, the image of the vertical line is not a straight line; rather it is projected as a curve with its convexity or concavity turned toward the fixation point, depending on whether the lens is convex or concave. However, this curve is seen with the peripheral retina and therefore, the subject does not necessarily perceive the deformation of the vertical line. The central static distortion has to be evaluated in the frontal plane by calculating the curvature of the projected image (A"B") formed after refraction in the periphery of the lens by rays originating from the straight line AB located in the frontal plane, at a given eccentricity (Figure 18).

*Figure 7*

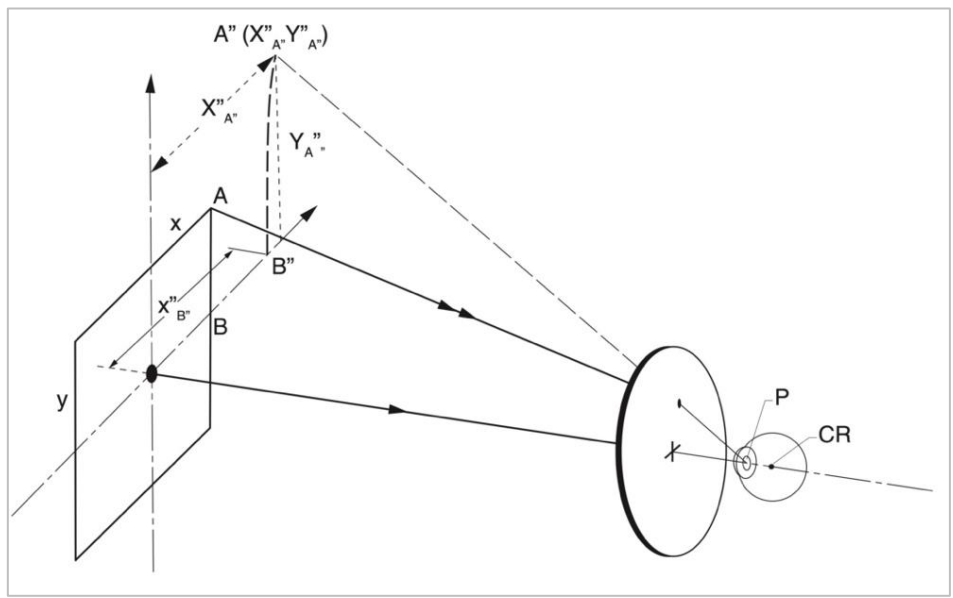

Central static distortion (Le Grand[11]; Fry[18]) occurs in the following conditions: the visual axis is superimposed with the optical axis of the lens and the fixation point, shown by a black spot, is at the center of the frontal plane (coordinates 0, 0). Point A is the upper right corner of a square with coordinates x, y in the frontal plane, this point is perceived in peripheral vision by the eye. At the center of the entrance pupil (P), the orientation of the refracted ray originating from A, gives the direction along which the image is projected in the frontal plane with coordinates $X"_{A"}$ and $Y"_{A"}$. In the presence of distortion, the projected image A"B" of the vertical line AB is not a straight line but a curve. As this projected line A"B" is seen in peripheral vision, its curvature is not necessarily perceived by the lens wearer. The horizontal component of central static distortion can be expressed by the sagitta of the curved line, the linear value ($X"_{A"} - x"_{B"}$) or by the ratio 100 [($X"_{A"} - x"_{B"}$) / ($x"_{B"} - x$)].

Secondly, dynamic distortion is that label given by Le Grand[11] to the distortion viewed when the subject rotates his eye from the primary position of gaze towards the edge of the lens to fix the foot B of a vertical line BA at the periphery of the frontal plane and then scans, in foveal vision, the image of the line BA seen through the periphery of the lens. In this case, the center of rotation of the eye (CR) is the reference point on which the visual axis lies when it explores the image of the vertical line (Figure 19). This is also the point on which the rays originating from the vertical line lie after being refracted by the periphery of the ophthalmic lens. The center of rotation of the eye is the reference point for finding the projection (B"A") of the image in the frontal plane (Figure 19).

*Figure 8*

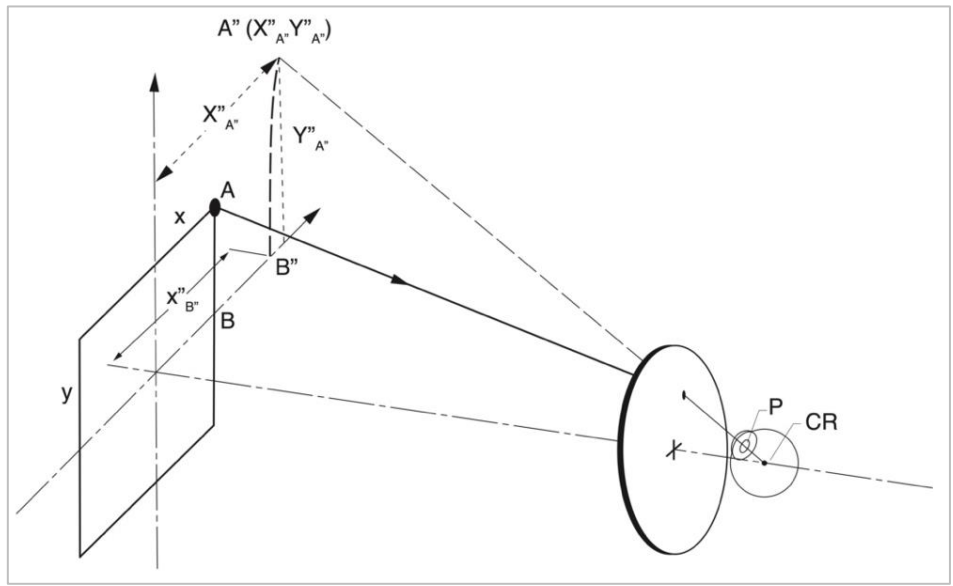

Dynamic distortion (Le Grand[11]), occurs when the eye moves behind the periphery of the lens and scans in foveal vision the projected image A"B". In this case, the orientation at the center of rotation (CR) of the refracted ray originating from point A (x, y) gives the direction along which the image of this point is projected (A") in the frontal plane. As the center of rotation (CR) rather than the center of the entrance pupil (P) is the reference point for the skew ray tracing, the various coordinates of the projected images will differ from the values observed with central static distortion.

Finally, Le Grand[11] identifies the lateral static distortion as the aberration inducing a deformation of the image observed in a secondary position of gaze when the subject fixates with his fovea the foot B of the vertical line BA situated in the periphery of the frontal plane (Figure 20) and perceives the curve of the image given by the periphery of the lens. In this position of gaze behind the lens, the subject's visual axis passes through the edge of the lens and is no more superimposed with the optical axis. The foot B of the vertical line is observed in foveal vision and the center of rotation of the eye (CR) is the point of reference aligned with the visual axis when the fixation point is B. However, the extremity A of the vertical line BA is viewed in peripheral vision and the center of the entrance pupil (P) is the point on which lie the luminous rays originating from A and from all points of the vertical line, except the ray coming from point B (Figure 20). The use of two different reference points (the center of rotation of the eye and the center of the entrance pupil) within the concept of lateral static distortion is certainly Le Grand's[11] primary contribution to the study of this aberration. Le Grand[11] concluded that lateral static distortion is the most disturbing distortion in practice since the perception of deformations implies foveal function without having to explore the vertical line in its entirety. In fact, as the subject must assess the deformation of a line passing through the fovea, he finds himself in the situations where hyperacuity is demonstrable. Therefore, lateral static distortion is more likely to be perceived.

*Figure 9*

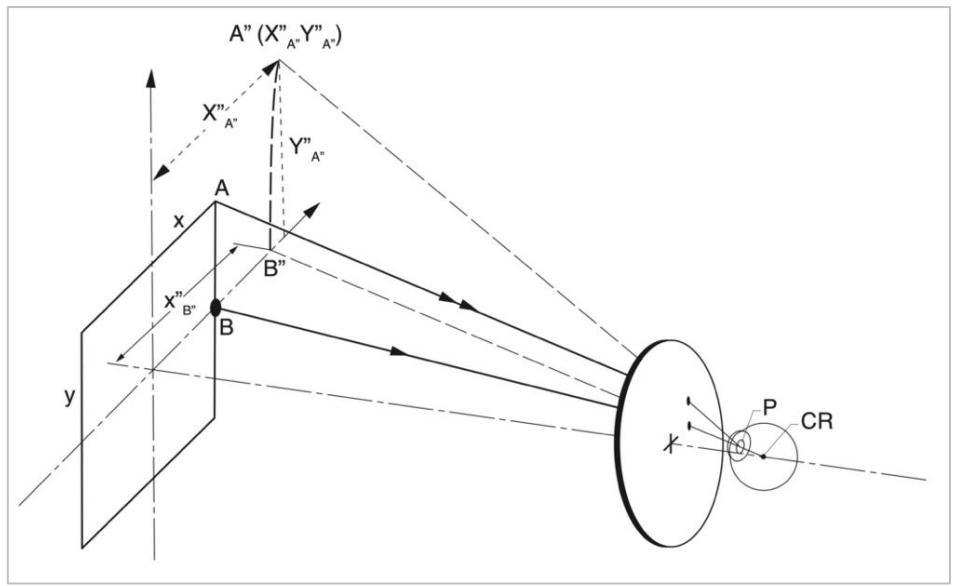

Lateral static distortion (Le Grand[11]) occurs when the visual axis is not superimposed with the optical axis of the lens during the fixation of point B. The direction along which the image of B is projected in the frontal plane (B") depends on the orientation of the refracted ray at the center of the rotation of the eye (CR). The direction along which the image of point A is projected in the frontal plane (A") depends on the orientation of the refracted ray at the center of the entrance pupil (P). Even though A" is seen in peripheral vision, the curvature of the projected image A"B" is analysed by foveal vision in an hyperacuity task, therefore lateral static distortion can be perceived and can be a disturbing aberration.

Le Grand[11] attempted to quantify the different types of distortions existing for lenses of powers of +6.00 and −10.00 D. However, the calculations he used for ray tracing induced him to determine the opposite of the distortion caused by a lens in a normal environment: he computed the curvature required for a line in the object space so that, seen through a lens, its image is a vertical line. Le Grand[11] made calculations for a line offset by 30 degree with respect to the optical axis of the lens. By comparing the values of the curvatures obtained by computation to the thresholds of detection of a curve established by Bourdon,[19] Le Grand[11] concluded that central static distortion was perceived minimally whereas dynamic and lateral distortion, were more likely to be noticed, and had almost similar values.

Independently from Le Grand[11], Fry[18] used a concept of distortion which corresponds to central static distortion and established the appropriate method to represent the deformation of images caused by this distortion. Fry[18] suggested the use of a skew ray tracing in order to find both the direction of an incident ray coming from a point at the periphery of the frontal plane, and the direction of the corresponding refracted ray aiming the center of the entrance pupil. According to Fry[18], the portion of the ray passing through the center of the entrance pupil represents the direction in which the image of the point formed by the periphery of the lens is viewed by the ametropic subject (Figure 18 and Figure 20). The projection in this direction helps to determine the intercept of this ray to the frontal plane and hence, the co-ordinates of the projected image in this reference plane. Therefore, it is possible to determine the deformation of the image as perceived by the lens wearer.

For a conventional lens of +6.00 D, Fry[18] put forward a representation of the deformation associated with central static distortion. For a given meridian, Fry[18] suggested that if s represents the distance between a given object point and the origin of the co-ordinates in the reference frontal plane, and s' represents the corresponding distance for the projected image, then the ratio ds'/ds considered as a function of s constitutes a means to represent distortion. Fry[18] was for the most part interested in central static distortion, and he did not apply the representation method he proposed to the other types of distortion defined by Le Grand.[11]

The value of each type of distortion can be determined by analyzing the deformation a square grid of known linear or angular dimensions when seen through a lens (Figure 21a and 10b). Simonet *et al.*[20] showed that the coordinates of the corresponding projected point in the reference plane can be determined from the coordinates of any given point in the object frontal plane using the skew ray tracing method. Figure 21a and 10b illustrate respectively central static distortion and lateral static distortion for a +5 D lens with a Tscherning design (refractive index is 1.61, approximate power of the front surface power is +10.75 D and lens diameter is 70 mm). The object square grid is located in the frontal plane at 5 m from the entrance pupil of the uncorrected subject. The center of the grid is aligned with the optical axis of the lens, the angular eccentricities of the vertical and horizontal lines at the margins of the grid are ± 30 degrees from the center of the grid. The skew ray tracing computation assumed a vertex distance of 14 mm, a value of 27 mm for the radius of the vertex sphere and a center of the entrance pupil (P) located at 10 mm in front of the center of rotation of the eye (CR).

*Figure 10*

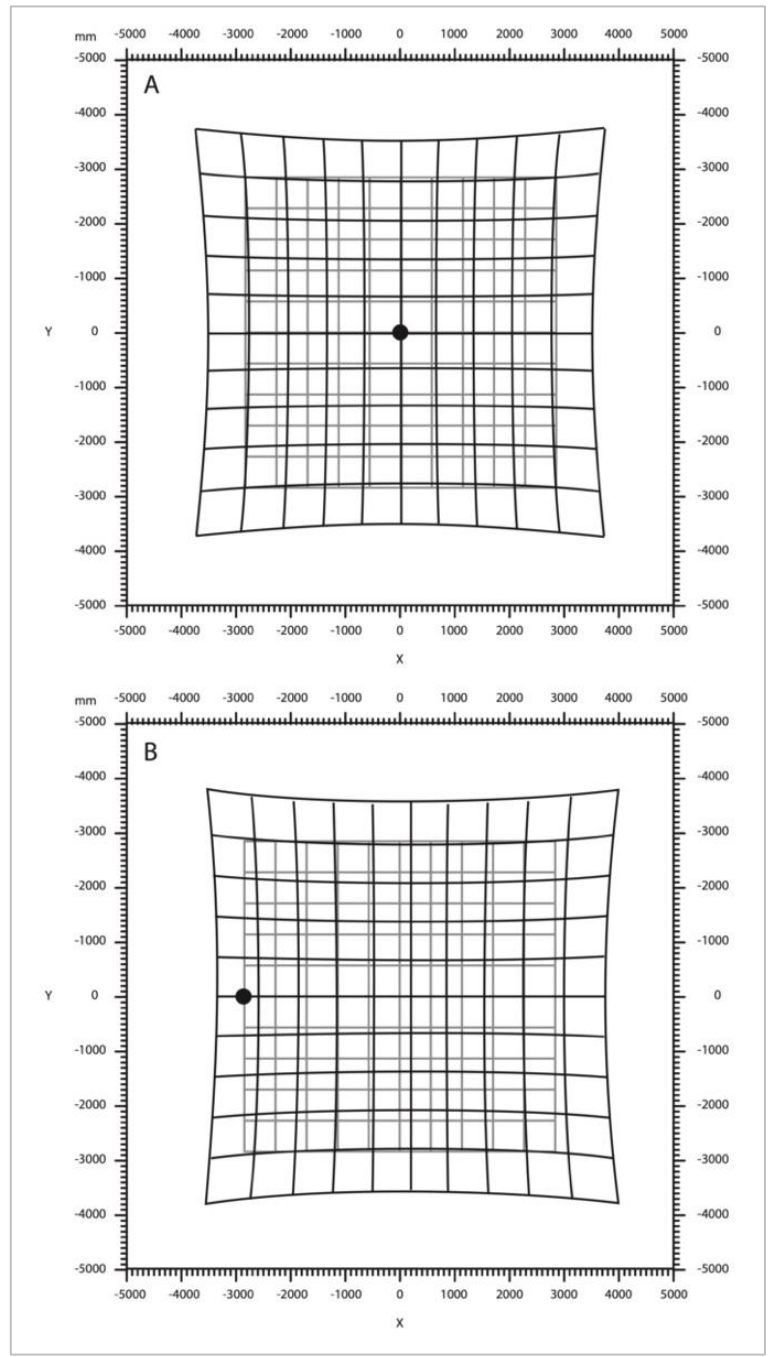

A - Central static distortion. B - Lateral static distortion. Both distortions are computed for a +5 D lens with a Tscherning design (n = 1.61, approximate power of the front surface = +10,75 D, diameter = 70 mm). The square object grid is represented by grey lines in the figure. The coordinates are reported in a frontal plane located at 5 m from the entrance pupil of the uncorrected subject. The center of the square grid is aligned with the optical axis of the lens, the angular eccentricities of the vertical and horizontal margins of the grid are ± 30 degrees from the center of the grid. The skew ray tracing computation assumed a vertex distance of 14 mm, a value of 27 mm for the radius of the vertex sphere and a center of the entrance pupil (P) located at 10 mm in front of the center of rotation of the eye (CR).

For an object square grid having an upper right corner which coordinates are x and y in the frontal plane, distortion will bend the margins of the projected grid such as X" and Y" will be the coordinates of the projected image of the upper right corner of the distorted grid, whereas x" will be the horizontal coordinate of the projected image of the middle of the vertical margin of the distorted grid and y" will be the vertical coordinate of the projected image of the middle of the horizontal margin.

Distortion could then be expressed by the sagitta of the curved image of the lateral margin of the grid.

$$D_{Horiz.} = (X"-x) - (x"-x) = (X"-x") \quad \text{and} \quad D_{Vert.} = (Y"-y) - (y"-y) = (Y"-y")$$

This sagitta can be presented as a linear measure or, better yet, as a value in visual angle which can be compared to hyperacuity thresholds. In the case of lateral static distortion, the reference margin of the grid is that fixated by the eye.

Distortion could also be expressed as a percentage by the ratio between the sagitta of the curved margin and the coordinate of the projected image of the middle of the lateral margin:

$$\text{Equation 6: } D\ (\%)_{Horiz.} = 100\ [(X" - x") / x"]$$
$$\text{Equation 7: } D\ (\%)_{Vert.} = 100\ [(Y" - y") / y"]$$

or, in an alternative approach, by the ratio between the sagitta of the curved margin and the distance separating the projected image of the middle of this margin (x") from the middle of the marginal side of the object square grid (x) in the frontal plane.

$$\text{Equation 8: } D(\%)_{Horiz.} = 100\ [(X" - x")\ /\ (x"-x)]$$

$$\text{Equation 9: } D(\%)_{Vert.} = 100\ [(Y" - y")\ /\ (y"-y)]$$

Because the two last equations consider the size of the object grid, they seem to be a better expression for quantifying distortion in order to compare the values of this aberration for different correcting power or different lens designs.

Le Grand's[11] concepts combined with Fry's[18] representation allow modeling of different types of distortions affecting the periphery of ophthalmic lenses. The display of central or lateral static distortions gives the deformation of the grid as it is presented to the lens wearer, whereas the display of dynamic distortion represents the variation of the actual prismatic effect occurring on the visual axis.

Simonet *et al.*[20] compared both types of static distortions for various corrective powers and different lens designs. They have established that, for a lens with a given design, lateral static distortion is slightly higher than central static distortion and that, for a given dioptric value, these two aberrations are of greater importance for plus lenses than minus lenses. Moreover, they calculated that both central static distortion and lateral static distortion could only be cancelled out on low negative powered lenses (<1.00 D) during the observation of a square grid presenting an eccentricity of ± 30 degrees in the visual field for its lateral margins. These lenses with spherical surface design can also be corrected for oblique astigmatism. Furthermore, Simonet *et al.* [20] showed that, for a given lens power, both static distortions increase as the lens base curve becomes flatter, but the use of aspheric surfaces reduces the static distortions occurring with flatter lenses.

Le Grand-Fry's approach for modeling distortion can be used to report the anamorphic distortion induced by abnormal pantoscopic angle or the deformation of an image due to a progressive power lens.[21] Figure 22 illustrates the distortion induced by a positive pantoscopic angle of 20 degrees, a value is at least twice that expected, for a +5 D lens where the optical center is located 4 mm below the center of the entrance pupil. In case of anamorphic distortion, the quantification of the aberration by a single parameter is not possible and the display of the complete distorted grid is necessary for the assessment of the perceptual effect.

Experimental measurements of distortion induced by ophthalmic lenses, other than afocal prisms, are scarce. Of these, the study by Fowler[22] reporting subjective evaluation of distortion through aphakic ophthalmic lenses seems to correspond to the measure of lateral static distortion or of dynamic distortion. The results demonstrate that, on aphakic lenses, this distortion can be easily observed and quantified.

*Figure 11*

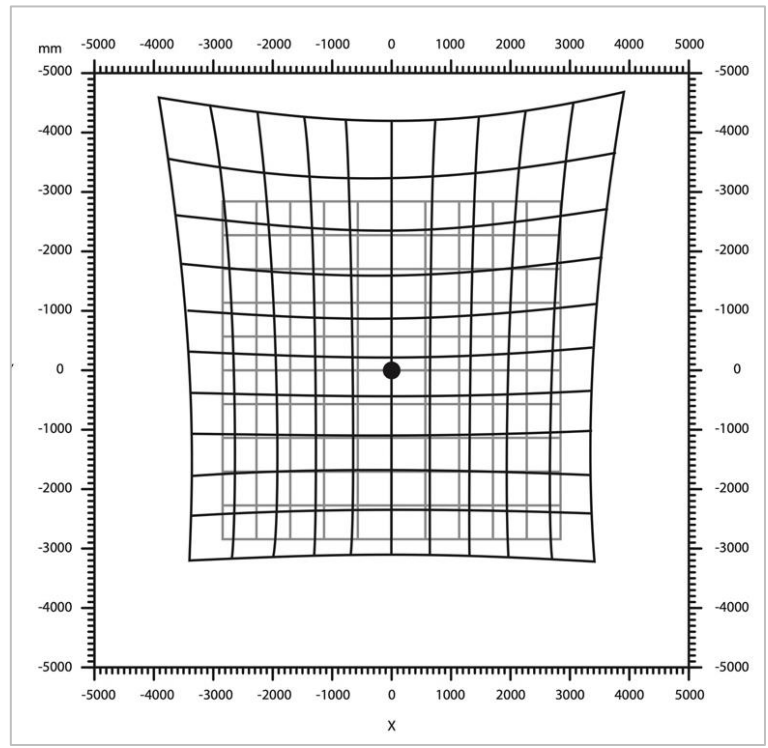

Anamorphic central static distortion induced for a +5 D lens by a pantoscopic angle of +20 degrees and by a location of the optical center of the lens 4 mm below the center of the entrance pupil. The characteristics of the lens, the size of the object square grid and the position of the frontal plane are similar to the conditions of figure 10.

## 6 Conclusion

Distortion, the aberration introduced by the periphery of an ophthalmic lens, is worthy of study in physiological optics only if it can be analyzed and quantified with respect to the perception of the lens wearer.

The concept initially developed by Tscherning[3, 4] has proved inappropriate because it is associated with only ocular rotation and seems more related to the problem of gauging the fixation field than distortion (Le Grand[7]). All subsequent similar approaches (Henker,[2] Emsley,[1, 10] Bennett[8]) associated with either field angle magnification or an ocular rotation factor present the same shortcoming and do not allow one to determine the exact deformation of the image of a straight line viewed through the periphery of the lens.

The approach which consists of applying the concept of distortion originating from instrumental optics is theoretically valid (Emsley,[1, 10] Jalie,[14]; Katz,[15] Atchison and Smith[16]); however, ordinary distortion, stationary distortion, peripheral distortion or rotatory distortion, either evaluated on the flat secondary focal plane or on the far-point sphere, remain of little interest if one seeks to analyze distortion in terms of the perception of the lens wearer. In fact, there is no physiological basis for the comparison between the position of two images, one paraxial and one resulting from an oblique gaze position, whether either the center of the entrance pupil or the center of rotation of the eye is considered as the reference point. The perception of the deformation of an image is obtained from observing a single image not from the simultaneous comparison of two images.

Finally, the perceptual approach to distortion developed by Le Grand[11] and by Fry[18] appears to be the preferred concept to analyze, quantify and model the deformation of images introduced by the periphery of ophthalmic lenses[23]. We recommend that a researcher or clinician who examines the distortion introduced by ophthalmic lenses should use a Le Grand-Fry approach.

## 7   Acknowledgments

Supported by the NSERC-Essilor Industrial Research Chair to Jocelyn Faubert and NSERC operating grants OGP0105658 to Pierre Simonet and OGP0121333 to Jocelyn Faubert. The authors thank Denis Latendresse for his technical help with the figures.

# 8  Figures captions

*Figure 12*

In the distortion linked to ocular rotation, the principal ray intersects the optical axis of the lens at the center of rotation of the eye (CR). The ocular rotation behind the lens is angle *u'* and angle *u* is the eccentricity of the fixation point M in the field of fixation.

*Figure 13*

Rotatory distortion, according to Emsley[1], is associated with the change in the angular extent of the field of fixation (*du*) linked to 1 degree ocular rotation (*du'*) while the eye moves from the center to the periphery of the ophthalmic lenses.

*Figure 14*

The ocular rotation factor W (Bennett and Edgar[12, 13]) is the ratio tan *u'* / tan $u_o$. M" is the apparent position of an object M seen through the lens after an ocular rotation *u'* while the object M is seen by the naked eye with an ocular rotation equal to $u_o$. According to these authors, distortion results from the fact that W is not constant in all directions of gaze.

*Figure 15*

In ordinary distortion Emsley[1]; Jalie[14]; Katz[15]) or in stationary distortion (Atchison and Smith[16]), the principal ray, parallel to the paraxial ray having an incidence *u*, intersects the optical axis of the lens, after refraction, at the center of the entrance pupil of the eye. Distortion is quantified by M'M1, the difference of position in the plane of the paraxial secondary focal point between the principal ray and the paraxial ray. This type of distortion corresponds to the classical approach in optical instrumentation.

*Figure 16*

In rotatory distortion, according to Jalie[14] and to Atchison and Smith[16], the principal ray, parallel to the paraxial ray having an incidence *u*, intersects the optical axis of the lens, after refraction, at the center of rotation of the eye (CR). Distortion is quantified by M'M1, the difference of position in the plane of the paraxial secondary focal point between the principal ray and the paraxial ray.

*Figure 17*

In rotatory distortion, according to Katz[15], the principal ray, parallel to the paraxial ray having an incidence *u*, intersects the optical axis of the lens, after refraction, at the center of rotation of the eye (CR). Distortion is quantified by M'M1, the difference of position on the far-point sphere between the principal ray and the paraxial ray.

*Figure 18*

Central static distortion (Le Grand[11]; Fry[18]) occurs in the following conditions: the visual axis is superimposed with the optical axis of the lens and the fixation point, shown by a black spot, is at the center of the frontal plane (coordinates 0, 0). Point A is the upper right corner of a square with coordinates x, y in the frontal plane, this point is perceived in peripheral vision by the eye. At the center of the entrance pupil (P), the orientation of the refracted ray originating from A, gives the direction along which the image is projected in the frontal plane with coordinates $X"_{A"}$ and $Y"_{A"}$. In the presence of distortion, the projected image A"B" of the vertical line AB is not a straight line but a curve. As this projected line A"B" is seen in peripheral vision, its curvature is not necessarily perceived by the lens wearer. The horizontal component of central static distortion can be expressed by the sagitta of the curved line, the linear value ($X"_{A"} - x"_{B"}$) or by the ratio $100 [(X"_{A"} - x"_{B"}) / (x"_{B"} - x)]$.

*Figure 19*

Dynamic distortion (Le Grand[11]), occurs when the eye moves behind the periphery of the lens and scans in foveal vision the projected image A"B". In this case, the orientation at the center of rotation (CR) of the refracted ray originating from point A (x, y) gives the direction along which the image of this point is projected (A") in the frontal plane. As the center of rotation (CR) rather than the center of the entrance pupil (P) is the reference point for the skew ray tracing, the various coordinates of the projected images will differ from the values observed with central static distortion.

*Figure 20*

Lateral static distortion (Le Grand[11]) occurs when the visual axis is not superimposed with the optical axis of the lens during the fixation of point B. The direction along which the image of B is projected in the frontal plane (B") depends on the orientation of the refracted ray at the center of the rotation of the eye (CR). The direction along which the image of point A is projected in the frontal plane (A") depends on the orientation of the refracted ray at the center of the entrance pupil (P). Even though A" is seen in peripheral vision, the curvature of the projected image A"B" is analysed by foveal vision in an hyperacuity task, therefore lateral static distortion can be perceived and can be a disturbing aberration.

*Figure 21*

A - Central static distortion. B - Lateral static distortion. Both distortions are computed for a +5 D lens with a Tscherning design (n = 1.61, approximate power of the front surface = +10,75 D, diameter = 70 mm). The square object grid is represented by grey lines in the figure. The coordinates are reported in a frontal plane located at 5 m from the entrance pupil of the uncorrected subject. The center of the square grid is aligned with the optical axis of the lens, the angular eccentricities of the vertical and horizontal margins of the grid are ± 30 degrees from the center of the grid. The skew ray tracing computation assumed a vertex distance of 14 mm, a value of 27 mm for the radius of the vertex sphere and a center of the entrance pupil (P) located at 10 mm in front of the center of rotation of the eye (CR).

*Figure 22*

Anamorphic central static distortion induced for a +5 D lens by a pantoscopic angle of +20 degrees and by a location of the optical center of the lens 4 mm below the center of the entrance pupil. The characteristics of the lens, the size of the object square grid and the position of the frontal plane are similar to the conditions of figure 10.